\documentclass[preprint,showpacs,preprintnumbers,amsmath,amssymb]{revtex4}

\usepackage{graphicx}% Include figure files
\usepackage{dcolumn}% Align table columns on decimal point
\usepackage{bm}% bold math

\newcommand{\Li}{\mathop{\mathrm{Li}}\nolimits}

\begin{document}

\preprint{DESY~06-074\hspace{11.5cm} ISSN 0418-9833}
\preprint{July 2006\hspace{14.9cm}}

\boldmath
\title{Strong-Coupling Constant with Flavor Thresholds at Five Loops in the
$\overline{\mathrm{MS}}$ Scheme}
\unboldmath

\author{B.A.~Kniehl}
\author{A.V.~Kotikov}
\altaffiliation[Also at ]{Bogolyubov Laboratory for Theoretical Physics, JINR,
141980 Dubna (Moscow region), Russia.}
\author{A.I.~Onishchenko}
\altaffiliation[Also at ]{Theoretical Physics Department, Petersburg Nuclear
Physics Institute, Orlova Roscha, 188300 Gatchina, Russia.}
\author{O.L.~Veretin}
\altaffiliation[Also at ]{Petrozavodsk State University, 185910 Petrozavodsk,
Karelia, Russia.}
\affiliation{{II.} Institut f\"ur Theoretische Physik, Universit\"at Hamburg,
22761 Hamburg, Germany}

%\date{\today}
\date{June 5, 2006}

\begin{abstract}
We present in analytic form the matching conditions for the strong-coupling
constant $\alpha_s^{(n_f)}(\mu)$ at the flavor thresholds to four loops in
the modified minimal-subtraction scheme.
Taking into account the present knowledge on the coefficient $\beta_4$ of the
Callan-Symanzik beta function of quantum chromo-dynamics, we thus derive a
five-loop formula for $\alpha_s^{(n_f)}(\mu)$ together with appropriate
relationships between the asymptotic scale parameters $\Lambda^{(n_f)}$ for
different numbers of flavors $n_f$.
\end{abstract}

\maketitle

The strong-coupling constant $\alpha_s^{(n_f)}(\mu)=g_s^2/(4\pi)$, where $g_s$
is the gauge coupling of quantum chromo-dynamics (QCD), is a fundamental
parameter of the standard model of elementary particle physics;
its value $\alpha_s^{(5)}(M_Z)$ is listed among the constants of nature
in the Review of Particle Physics \cite{pdg}.
Here, $\mu$ is the renormalization scale, and $n_f$ is the number of active
quark flavors $q$, with mass $m_q\ll\mu$.
The $\mu$ dependence of $\alpha_s^{(n_f)}(\mu)$ is controlled by the
Callan-Symanzik beta function of QCD,
\begin{eqnarray}
\label{rge}
\mu^2\frac{d}{d\mu^2}\,\frac{\alpha_s^{(n_f)}(\mu)}{\pi}
&=&\beta^{(n_f)}\left(\frac{\alpha_s^{(n_f)}(\mu)}{\pi}\right)
\nonumber\\
&=&{}-\sum_{N=0}^\infty\beta_N^{(n_f)}
\left(\frac{\alpha_s^{(n_f)}(\mu)}{\pi}\right)^{N+2}.
\end{eqnarray}
The calculation of the one-loop coefficient $\beta_0^{(n_f)}$ about 33 years
ago \cite{gro} has led to the discovery of asymptotic freedom and to the
establishment of QCD as the theory of strong interactions, an achievement that
was awarded by the 2004 Nobel Prize in Physics.
In the class of schemes where the beta function is mass independent, which
includes the minimal-subtraction (MS) schemes of dimensional regularization
\cite{bol}, $\beta_0^{(n_f)}$ and $\beta_1^{(n_f)}$ \cite{jon} are universal.
The results for $\beta_2^{(n_f)}$ \cite{tar} and $\beta_3^{(n_f)}$ \cite{rit}
are available in the modified MS ($\overline{\rm MS}$) scheme \cite{bar}.
As for $\beta_4^{(n_f)}$, the term proportional to $n_f^4$,
\begin{equation}
\beta_4^{(n_f)}=\left[\frac{1205}{2985984}-\frac{19}{10368}\zeta(3)\right]
n_f^4+O\left(n_f^3\right),
\end{equation}
where $\zeta$ is Riemann's zeta function, was found in the large-$n_f$
expansion \cite{Gracey:1996up}, while the residual terms, of 
$O(n_f^3)$ and below, are presently unknown.
However, the latter were estimated by an educated guess, through weighted
asymptotic Pad\'e approximant predictions (WAPAP's), which are improved by
including asymptotic corrections with respect to the usual Pad\'e approximants
and performing a weighted average over negative values of $n_f$
\cite{Ellis:1997sb}.
In the case of $\beta_3^{(n_f)}$, leaving aside the quartic Casimir terms,
which appear there for the first time, the WAPAP's approximate the exact
coefficients of $n_f^n$ with $n=0,1,2$ amazingly well, at the one-percent
level.
One may thus expect that the WAPAP's for $\beta_4^{(n_f)}$ work similarly
well, except for the quartic Casimir terms, which cannot be predicted quite
as reliably.
For the reader's convenience, $\beta_N^{(n_f)}$ $(N=0,\ldots,4)$ are listed
for the $n_f$ values of practical interest in Table~\ref{tab:beta}.
\renewcommand{\arraystretch}{1.2}
\begin{table}
\caption{\label{tab:beta}$\overline{\rm MS}$ values of $\beta_N^{(n_f)}$ for
variable $n_f$.
$\beta_4^{(n_f)}$ is estimated by WAPAP's with quartic Casimir terms omitted
\cite{Ellis:1997sb}.}
\begin{ruledtabular}
\begin{tabular}{cccccc}
$n_f$ & $\beta_0^{(n_f)}$ & $\beta_1^{(n_f)}$ & $\beta_2^{(n_f)}$ &
$\beta_3^{(n_f)}$ & $\beta_4^{(n_f)}$\\
\hline
$3$
& $\frac{9}{4}$
& $4$
& $\frac{3863}{384}$
& $\frac{445}{32}\zeta(3)+\frac{140599}{4608}$
& 162
\\
$4$
& $\frac{25}{12}$
& $\frac{77}{24}$
& $\frac{21943}{3456}$
& $\frac{78535}{5184}\zeta(3)+\frac{4918247}{373248}$
& 119
\\
$5$
& $\frac{23}{12}$
& $\frac{29}{12}$
& $\frac{9769}{3456}$
& $\frac{11027}{648}\zeta(3)-\frac{598391}{373248}$
& 107
\\
$6$
& $\frac{7}{4}$
& $\frac{13}{8}$
& $-\frac{65}{128}$
& $\frac{11237}{576}\zeta(3)-\frac{63559}{4608}$
& 124
\\
\end{tabular}
\end{ruledtabular}
\end{table}
\renewcommand{\arraystretch}{1}

In MS-like renormalization schemes, the Appelquist-Carazzone decoupling
theorem \cite{app} does not in general apply to quantities that do not
represent physical observables, such as beta functions or coupling constants, 
{\it i.e.}, quarks with mass $m_q\gg\mu$ do not automatically decouple.
The standard procedure to circumvent this problem is to render decoupling
explicit by using the language of effective field theory.
As an idealized situation, consider QCD with $n_l=n_f-1$ massless quark
flavors and one heavy flavor $h$, with mass $m_h\gg\mu$.
Then, one constructs an effective $n_l$-flavor theory by requiring 
consistency with the full $n_f$-flavor theory at the heavy-quark threshold
$\bar\mu={\cal O}(m_h)$.
This leads to a nontrivial matching condition between the couplings of the two
theories.
Although, $\alpha_s^{(n_l)}(m_h)=\alpha_s^{(n_f)}(m_h)$ at leading and
next-to-leading orders, this relationship does not generally hold at higher
orders in the $\overline{\rm MS}$ scheme, {\it i.e.},
$\alpha_s^{(n_f)}(\mu)$ starts to exhibit finite discontinuities at the flavor
thresholds.
If the $\mu$ evolution of $\alpha_s^{(n_f)}(\mu)$ is to be performed at $N+1$
loops, {\it i.e.}, with the highest coefficient in Eq.~(\ref{rge}) being
$\beta_N^{(n_f)}$, then consistency requires that the matching conditions be
implemented in terms of $N$-loop formulae.
Then, the residual $\mu$ dependence of physical observables will be of order
$N+2$.
The QCD matching conditions at the flavor thresholds to two \cite{ber} and
three \cite{Chetyrkin:1997sg} loops are known in analytical form; they are
routinely used in the literature and even copied to the Review of Particle
Physics \cite{pdg}.
%\cite{Caso:1998tx}.
Recently, the four-loop result was found, in semi-analytical form
\cite{Schroder:2005hy}.
In fact, the most intricate four-loop tadpole master integrals involving one
non-vanishing mass among the basic set that enters any such calculation could
so far only be computed numerically, with limited precision
\cite{Schroder:2005hy,Schroder:2005va,Chetyrkin:2006xg}.
It is the purpose of this Letter, to overcome this bottle-neck by presenting
the four-loop matching condition for $\alpha_s^{(n_f)}(\mu)$ entirely in terms
of elementary transcendental numbers.
This requires the analytic evaluation of the massive four-loop tadpole diagram
that is called $X_0$ or $T_{91}$ in the recent literature
\cite{Schroder:2005hy}.
Together with the results of Ref.~\cite{Kniehl:2005yc}, we thus enhance the
knowledge of the basic set of massive four-loop tadpole master integrals in
analytic form.

Prior to explaining the core of this analysis and presenting our analytic
result for the four-loop matching condition for $\alpha_s^{(n_f)}(\mu)$, we
derive the five-loop formula for this coupling for fixed value of $n_f$.
In order to simplify the notation, we introduce the couplant
$a^{(n_f)}(\mu)=\alpha_s^{(n_f)}(\mu)/\pi$ and omit the labels $\mu$ and $n_f$
wherever confusion is impossible.
Integrating Eq.~(\ref{rge}) leads to
\begin{eqnarray}
\label{con}
\lefteqn{\ln\frac{\mu^2}{\Lambda^2}=\int\frac{da}{\beta(a)}
=\frac{1}{\beta_0}\left[\frac{1}{a}+b_1\ln a
+a\left(-b_1^2+b_2\right)
\right.}
\nonumber\\
&&{}+a^2\left(\frac{b_1^3}{2}-b_1b_2+\frac{b_3}{2}\right)
+a^3\left(-\frac{b_1^4}{3}+b_1^2 b_2-\frac{b_2^2}{3}
\right.\nonumber\\
&&{}-\left.\left.
\frac{2}{3}b_1b_3+\frac{b_4}{3}
\right)+O(a^4)\right]+C,
\end{eqnarray}
where $b_N=\beta_N/\beta_0$ ($N=1,\ldots,4$), $\Lambda$ is the so-called
asymptotic scale parameter, and $C$ is an arbitrary constant.
The second equality in Eq.~(\ref{con}) is obtained by expanding the integrand.
The conventional $\overline{\rm MS}$ definition of $\Lambda$, which we adopt,
corresponds to choosing $C=(b_1/\beta_0)\ln\beta_0$ \cite{bar,fur}.
Iteratively solving Eq.~(\ref{con}) yields,
with $L=\ln(\mu^2/\Lambda^2)$,
\begin{eqnarray}
\label{alp}
\lefteqn{a=\frac{1}{\beta_0L}-\frac{b_1\ln L}{(\beta_0L)^2}
+\frac{1}{(\beta_0L)^3}\left[b_1^2(\ln^2L-\ln L-1)+b_2\right]}
\nonumber\\
&&{}+\frac{1}{(\beta_0L)^4}\left[
b_1^3\left(-\ln^3L+\frac{5}{2}\ln^2L+2\ln L-\frac{1}{2}\right)
\right.
\nonumber\\
&&{}-\left.
3b_1b_2\ln L+\frac{b_3}{2}\right]
+\frac{1}{(\beta_0L)^5}\left[b_1^4\left(\ln^4L-\frac{13}{3}\ln^3L
\right.\right.
\nonumber\\
&&-\left.
\frac{3}{2}\ln^2L+4\ln L+\frac{7}{6}\right)
+3b_1^2b_2(2\ln^2L-\ln L-1)
\nonumber\\
&&{}-b_1b_3\left(2\ln L+\frac{1}{6}\right)
+\left.\frac{5}{3}b_2^2+\frac{b_4}{3}\right]
+O\left(\frac{1}{L^6}\right).
\end{eqnarray}
The particular choice of $C$ \cite{bar,fur} in Eq.~(\ref{con}) is predicated
on the grounds that it suppresses the appearance of a term proportional to
$({\rm const.}/L^2)$ in Eq.~(\ref{alp}).

\begin{figure}
\begin{center}
\includegraphics[width=6cm]{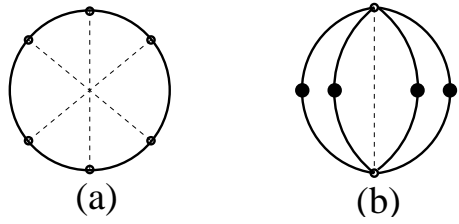}
\caption{\label{fig:dia}Four-loop tadpole diagrams (a) $X_0$ and (b) $J_0$.
Dashed and solid lines represent massless and massive propagators; a dot on
a line duplicates that propagator.}
\end{center}
\end{figure}
We now turn to the analytic evaluation of the four-loop matching condition for
$\alpha_s^{(n_f)}(\mu)$ at the flavor thresholds.
The underlying formalism was comprehensively explained in
Refs.~\cite{ber,Chetyrkin:1997sg}, and most of the technical issues related to
its application at four loops were already discussed in
Ref.~\cite{Schroder:2005hy}.
For lack of space, we thus concentrate here on the missing link of this
analysis beyond the scope of Ref.~\cite{Schroder:2005hy}, namely the analytic
evaluation of the massive four-loop tadpole diagram $X_0$, which is depicted
in Fig.~\ref{fig:dia}(a).
This task may be simplified by noticing that $X_0$ does not represent a master
integral, but may be reduced to simpler integrals with less lines, all of
which are analytically known \cite{Schroder:2005va}, some for a short time
only \cite{Kniehl:2005yc}, except for the one ($J_0$) shown in
Fig.~\ref{fig:dia}(b).
The integral $J_0$ is finite, and the coefficients of its expansion in
$\epsilon$, where $D=4-2\epsilon$ is the dimensionality of space-time, have
only one level of trancendentality \cite{Fleischer:1998nb}, {\it i.e.}, they
contain poly-logarithms $\Li_k$ and zeta functions $\zeta(k)$ with the same
value of $k$.
These properties reduce the number of terms and thus simplify the calculation.
In order to evaluate $J_0$, we temporarily introduce an artificial mass
splitting among the four massive lines in Fig.~\ref{fig:dia}(b), by assigning
the mass $m$ to any two of them and the mass $M$ to the other two.
We then perform an expansion in the ratio $x=m^2/M^2$ using the large-mass
expansion technique and recover the complete series in $x$ as explained in
Ref.~\cite{Fleischer:1998nb}.
Through $O(\epsilon^2)$, we have
\begin{eqnarray}
\lefteqn{(1-\epsilon)(m^2M^2)^{2\epsilon}m^2J_{0}
=\sum_{n=1}^{\infty}x^n\left\{-\frac{8}{(2n-1)^3} 
\right.}
\nonumber\\
&&{}+16\epsilon\left[
-\frac{\zeta(3)-\tilde{S}_3}{2n-1}+\frac{2 \tilde{S}_2}{(2n-1)^2}
+\frac{4 \tilde{S}_1}{(2n-1)^3}\right]
\nonumber\\
&&{}+16\epsilon^2\left[
\frac{9\zeta(4)+8\zeta(3)\tilde{S}_1-4\tilde{S}_2^2-8\tilde{S}_1\tilde{S}_3
-3\tilde{S}_4}{2n-1}
\right.
\nonumber\\
&&{}+\frac{4\zeta(3)-16\tilde{S}_1 \tilde{S}_2- 4 \tilde{S}_3}{(2n-1)^2}
-\frac{\zeta(2)+16\tilde{S}_1^2+4\tilde{S}_2}{(2n-1)^3}
\nonumber\\
&&{}-\left.\left.
\frac{2}{(2n-1)^5}\right]+O(\epsilon^3)\right\},
\label{J1b}
\end{eqnarray}
where we introduced the short-hand notation
$\tilde{S}_a =2^{a-2} S_a(2n-1) - S_a(n-1)$, with
$S_a(n)=\sum_{j=1}^n j^{-a}$ being harmonic sums, and omitted irrelevant terms
involving $\ln x$.
We then put $x=1$ and exploit the identities
\begin{eqnarray}
 \sum_{n=1}^{\infty}
\frac{ \tilde{S}_a}{(2n-1)^c} &=& -\frac{2^a}{8}
 \sum_{l=1}^{\infty}
\frac{1-(-1)^l}{l^c}[S_a(l)+2S_{-a}(l)],
\nonumber\\
 \sum_{n=1}^{\infty} 
\frac{ \tilde{S}_a \tilde{S}_b}{(2n-1)^c} &=& \frac{2^{a+b}}{32}
 \sum_{l=1}^{\infty} 
\frac{1-(-1)^l}{l^c}[S_a(l)+2S_{-a}(l)]
\nonumber\\
&&{}\times[S_b(l)+2S_{-b}(l)],
\end{eqnarray}
%where $l=2n-1$, 
at trancendentality levels $k=a+c$ and $k=a+b+c$,
respectively.
The sums with $k<5$ may be found in Ref.~\cite{Fleischer:1998nb}, while those
with $k=5$ may be obtained from there through integration,
\begin{equation}
  \sum_{l=1}^{\infty} 
\frac{f(l) x^l}{l^{c+1}} = \int^x_0 \frac{dy}{y}
  \sum_{l=1}^{\infty}
\frac{f(l) y^l}{l^c},
\end{equation}
where $f(l)=S_{\pm a}(l),\ldots$.
After some algebra, we find an analytic expression for $J_0$ and hence also
for $X_0$,
\begin{equation}
X_0=-318\zeta(4)\ln 2+\frac{873}{2}\zeta(5)-48 b_5+ O(\epsilon).
\end{equation}
Here and in the following, we use the constants
\begin{eqnarray}
a_4&=&\left(-2\zeta(2)+\frac{\ln^2 2}{3}\right)\ln^2 2
+8\Li_4\left(\frac{1}{2}\right),
\nonumber\\
a_5 &=&\frac{1}{3}\left(2\zeta(2)-\frac{\ln^2 2}{5}\right)\ln^3 2
+8\Li_5\left(\frac{1}{2}\right).
\end{eqnarray}

If we measure the matching scale $\bar\mu$ in units of the
$\overline{\rm MS}$ mass $m_h\left(\bar\mu\right)$, our result for the
ratio of $a^\prime=a^{(n_l)}\left(\bar\mu\right)$ to
$a=a^{(n_f)}\left(\bar\mu\right)$ reads
%\begin{widetext}
\begin{eqnarray}
\label{msb}
\lefteqn{\frac{a^\prime}{a}=
1
-a\frac{\ell}{6}
+a^2\left(\frac{\ell^2}{36}-\frac{11}{24}\ell+c_2\right)
+a^3\left[-\frac{\ell^3}{216}
\right.}
\nonumber\\
&&{}+\left.\ell^2\left(\frac{53}{576}-\frac{n_l}{36}\right)
+\ell\left(-\frac{955}{576}+\frac{67}{576}n_l\right)
+c_3\right]
\nonumber\\
&&{}+a^4
\left\{\frac{\ell^4}{1296}
+\ell^3\left(-\frac{1883}{10368}-\frac{127}{5184}n_l+\frac{n_l^2}{324}\right)
\right.
\nonumber\\
&&{}
+\ell^2\left(\frac{2177}{3456}-\frac{1483}{10368}n_l-\frac{77}{20736}n_l^2
\right)
+\ell\left[\frac{7391699}{746496} 
\right.
\nonumber\\
&&{}-\frac{2529743}{165888}\zeta(3)
+n_l\left(-\frac{110341}{373248}+\frac{110779}{82944}\zeta(3)\right)
\nonumber\\
&&{}+\left.\left.
\frac{6865}{186624}n_l^2\right]
+c_4
\right\}
+O(a^5),
\end{eqnarray}
%\end{widetext}
where $\ell=\ln\left[\bar\mu^2/m_h^2\left(\bar\mu\right)\right]$
and
\begin{eqnarray}
\label{cms}
&&c_2=\frac{11}{72},\qquad
c_3=\frac{564731}{124416}-\frac{82043}{27648}\zeta(3)
-\frac{2633}{31104}n_l,
\nonumber\\
&&c_4=\frac{291716893}{6123600}
-\frac{2362581983}{87091200}\zeta(3)
-\frac{76940219}{2177280}\zeta(4)
\nonumber\\
&&{}+\frac{9318467}{362880}\zeta(4)\ln2
-\frac{12057583}{483840}\zeta(5)
+\frac{3031309}{435456}a_4
\nonumber\\
&&{}+\frac{340853}{90720}a_5 
+n_l\left(
-\frac{4770941}{2239488} 
+\frac{3645913}{995328}\zeta(3) 
\right.
\nonumber\\
&&{}-\left.\frac{541549}{165888}\zeta(4)
+\frac{115}{576}\zeta(5)
+\frac{685}{41472}a_4\right)
\nonumber\\
&&{}+n_l^2\left(-\frac{271883}{4478976} 
+\frac{167}{5184}\zeta(3)\right).
\end{eqnarray}
The counterpart of Eq.~(\ref{msb}) in the on-shell scheme of mass
renormalization may be obtained by substituting the three-loop relationship
between $m_h(\mu)$ and the pole mass $M_h$ \cite{Melnikov:2000qh}.

\begin{figure}
\begin{center}
\includegraphics[width=8cm]{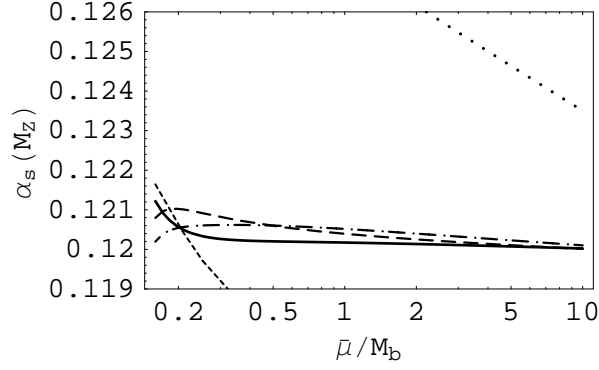}
\caption{\label{fig:asmz}$\bar\mu$ dependence of $\alpha_s^{(5)}(M_Z)$ from
$N$-loop evolution and $(N-1)$-loop matching, with $N=1$ (dotted), 2
(short-dashed), 3 (dot-dashed), 4 (long-dashed), and 5 (solid).}
\end{center}
\end{figure}
Going to higher orders, one expects, on general grounds, that the relationship
between $\alpha_s^{(n_l)}(\mu^\prime)$ and $\alpha_s^{(n_f)}(\mu)$, where
$\mu^\prime\ll\bar\mu\ll\mu$, becomes insensitive to the choice of
$\bar\mu$ as long as $\bar\mu={\cal O}(m_h)$.
This has been checked in Ref.~\cite{Chetyrkin:1997sg} for four-loop evolution
in connection with three-loop matching.
Armed with our new results, we are in a position to explore the situation at 
the next order.
As an example, we consider the crossing of the bottom-quark threshold.
In particular, we wish to study how the $\bar\mu$ dependence of the
relationship between $\alpha_s^{(4)}(M_\tau)$ and $\alpha_s^{(5)}(M_Z)$ is
reduced as we implement five-loop evolution with four-loop matching.
Our procedure is as follows.
We first calculate $\alpha_s^{(4)}(\bar\mu)$ with Eq.~(\ref{alp}) by
imposing the condition $\alpha_s^{(4)}(M_\tau)=0.34$ \cite{pdg}, then obtain
$\alpha_s^{(5)}(\bar\mu)$ from the on-shell version of Eq.~(\ref{msb})
with $M_b=4.85$~GeV \cite{pdg}, and finally compute $\alpha_s^{(5)}(M_Z)$ with
Eq.~(\ref{alp}).
For consistency, $N$-loop evolution must be accompanied by $(N-1)$-loop 
matching, {\it i.e.}, if we omit terms of ${\cal O}(1/L^{N+1})$ in
Eq.~(\ref{alp}), we need to discard those of ${\cal O}(a^N)$ in
Eq.~(\ref{msb}) at the same time.
In Fig.~\ref{fig:asmz}, the variation of $\alpha_s^{(5)}(M_Z)$ with
$\bar\mu/M_b$ is displayed for the various levels of accuracy, ranging
from one-loop to five-loop evolution.
For illustration, $\bar\mu$ is varied rather extremely, by almost two
orders of magnitude.
While the leading-order result exhibits a strong logarithmic behavior, the
analysis is gradually getting more stable as we go to higher orders.
The five-loop curve is almost flat.
Besides the $\bar\mu$ dependence of $\alpha_s^{(5)}(M_Z)$, also its
absolute normalization is significantly affected by the higher orders.
At the central scale $\bar\mu=M_b$, we encounter an alternating convergence
behavior.

As we have learned from Fig.~\ref{fig:asmz}, in higher orders, the actual
value of $\bar\mu$ does not matter as long as it is comparable to the
heavy-quark mass.
In the context of Eq.~(\ref{msb}), the choice $\bar\mu=\mu_h$, where
$\mu_h=m_h(\mu_h)$ is the renormalization-group (RG) invariant
$\overline{\rm MS}$ mass, is particularly convenient, since it eliminates the
RG logarithm $\ell$.
With this convention, we obtain from Eqs.~(\ref{con}), (\ref{alp}), and
(\ref{msb}) a simple relationship between $\Lambda^\prime=\Lambda^{(n_l)}$ and
$\Lambda=\Lambda^{(n_f)}$, viz
\begin{eqnarray}
\label{lam}
\lefteqn{\beta_0^\prime\ln\frac{\Lambda^{\prime2}}{\Lambda^2}=
\left(\beta_0^\prime-\beta_0\right)l+\left(b_1^\prime-b_1\right)\ln l
-b_1^\prime\ln\frac{\beta_0^\prime}{\beta_0}}
\\
&&{}+\frac{1}{\beta_0l}\left[b_1\left(b_1^\prime-b_1\right)\ln l
+b_1^{\prime2}-b_1^2-b_2^\prime+b_2+c_2\right]
\nonumber\\
&&{}+\frac{1}{(\beta_0l)^2}\left[
\frac{b_1^3}{2}(\ln^2l-1)
-b_1^\prime b_1^2\left(\frac{1}{2}\ln^2l-\ln l-1\right)
\right.
\nonumber\\
&&{}-b_1\left(b_1^{\prime2}-b_2^\prime+b_2+c_2\right)\ln l
-\frac{b_1^{\prime3}}{2}
+b_1^\prime\left(b_2^\prime-b_2-c_2\right)
\nonumber\\
&&{}-\left.\frac{1}{2}\left(b_3^\prime-b_3\right)+c_3\right]
+\frac{1}{(\beta_0l)^3}
\left\{-b_1^4\left(\frac{1}{3}\ln^3l-\frac{1}{2}\ln^2l
\right.\right.
\nonumber\\
&&{}-\left.\ln l-\frac{1}{6}\right)
+b_1^\prime b_1^3\left(\frac{1}{3}\ln^3l-\frac{3}{2}\ln^2l-\ln l
+\frac{1}{2}\right)
+b_1^2
\nonumber\\
&&{}\times\left(b_1^{\prime2}-b_2^\prime+b_2+c_2\right)(\ln^2l-\ln l-1)
+b_1\left[b_1^{\prime3}-2b_1^\prime
\right.
\nonumber\\
&&{}\times\left.\left(b_2^\prime-b_2-c_2\right)
+b_3^\prime-b_3-2c_3\right]\ln l
+\frac{b_1^{\prime4}}{3}
-b_1^{\prime2}\left(b_2^\prime
\right.
\nonumber\\
&&{}-\left.b_2-c_2\right)
+\left(b_2^\prime-b_2\right)
\left(\frac{b_2^\prime}{3}-\frac{2}{3}b_2-c_2\right)
-c_2^2
+b_1^\prime\left(\frac{2}{3}\right.
\nonumber\\
&&{}\times\left.
\left.b_3^\prime-\frac{b_3}{2}-c_3\!\right)\!
-\frac{b_1b_3}{6}
-\frac{1}{3}\left(b_4^\prime-b_4\right)
+c_4\!\right\}\!+O\!\left(\!\frac{1}{l^4}\!\right)\!,
\nonumber
\end{eqnarray}
where $l=\ln(\mu_h^2/\Lambda^2)$.
The ${\cal O}(1/l^3)$ term of Eq.~(\ref{lam}) is new.
Equation~(\ref{lam}) represents a closed four-loop formula for
$\Lambda^{(n_l)}$ in terms of $\Lambda^{(n_f)}$ and $\mu_h$.
For consistency, it should be used in connection with the five-loop
expression~(\ref{alp}) for $\alpha_s^{(n_f)}(\mu)$ with the understanding that
the underlying flavor thresholds are fixed at $\bar\mu=\mu_h$.
The inverse relation that gives $\Lambda^{(n_f)}$ as a function of
$\Lambda^{(n_l)}$ and $\mu_h$ emerges from Eq.~(\ref{lam}) via the 
substitutions $\Lambda\leftrightarrow\Lambda^\prime$;
$\beta_N\leftrightarrow\beta_N^\prime$ for $N=0,\ldots,4$; and $c_N\to-c_N$
for $N=2,3,4$.

In conclusion, we have extended the standard description of the
strong-coupling constant in the $\overline{\rm MS}$ renormalization scheme to
include five-loop evolution and four-loop matching at the flavor thresholds.
These results will be indispensable in order to relate the QCD predictions for
different observables at next-to-next-to-next-to-next-to-leading order.

\begin{acknowledgments}
We thank J.A.M.~Vermaseren for a useful communication regarding
Ref.~\cite{Ellis:1997sb}.
A.V.K. was supported in part by the RFBR Foundation through Grant No.\
05-02-17645-a and by the Heisenberg-Landau-Programm.
This work was supported in part by BMBF Grant No.\ 05 HT4GUA/4 and HGF Grant
No.\ NG-VH-008.
\end{acknowledgments}


\begin{thebibliography}{99}

\bibitem{pdg} Particle Data Group, S.~Eidelman {\it et al.},
Phys.\ Lett.\ B {\bf 592}, 1 (2004); URL: http://pdg.lbl.gov/.
%and 2005 partial update for the 2006
%edition available on the PDG WWW pages (URL: http://pdg.lbl.gov/).

\bibitem{gro} D.J.~Gross and F.~Wilczek,
Phys.\ Rev.\ Lett.\ {\bf30}, 1343 (1973);
H.D.~Politzer, {\it ibid.}\ {\bf30}, 1346 (1973).

\bibitem{bol} C.G.~Bollini and J.J.~Giambiagi,
Phys.\ Lett.\ {\bf40B}, 566 (1972);
G. 't~Hooft and M.~Veltman, Nucl.\ Phys.\ {\bf B44}, 189 (1972);
G. 't~Hooft, {\it ibid.}\ {\bf B61}, 455 (1973).

\bibitem{jon} D.R.T.~Jones, Nucl.\ Phys.\ {\bf B75}, 531 (1974);
W.E.~Caswell, Phys.\ Rev.\ Lett.\ {\bf33}, 244 (1974);
\'E.Sh.\ Egoryan and O.V.~Tarasov, Teor.\ Mat.\ Fiz.\ {\bf41}, 26 (1979)
[Theor.\ Math.\ Phys.\ {\bf41}, 863 (1979)].

\bibitem{tar} O.V.~Tarasov, A.A.~Vladimirov, and A.Yu.\ Zharkov,
Phys.\ Lett.\ {\bf93B}, 429 (1980);
S.A.~Larin and J.A.M.~Vermaseren, {\it ibid.}\ B {\bf303}, 334 (1993).

\bibitem{rit} T.~van Ritbergen, J.A.M.~Vermaseren, and S.A.~Larin,
Phys.\ Lett.\ B {\bf400}, 379 (1997).

\bibitem{bar} W.A.~Bardeen, A.J.~Buras, D.W.~Duke, and T.~Muta,
Phys.\ Rev.\ D {\bf18}, 3998 (1978).

\bibitem{Gracey:1996up}
J.A.~Gracey,
Phys.\ Lett.\ B {\bf373}, 178 (1996).

\bibitem{Ellis:1997sb}
J.~Ellis, I.~Jack, D.R.T.~Jones, M.~Karliner, and M.A.~Samuel,
Phys.\ Rev.\ D {\bf57}, 2665 (1998).

\bibitem{app} T.~Appelquist and J.~Carazzone,
Phys.\ Rev.\ D {\bf11}, 2856 (1975).

\bibitem{ber}
W.~Bernreuther and W.~Wetzel,
Nucl.\ Phys.\ {\bf B197}, 228 (1982); {\bf B513}, 758(E) (1998);
S.A.~Larin, T.~van Ritbergen, and J.A.M.~Vermaseren,
{\it ibid.}\ {\bf B438}, 278 (1995).

\bibitem{Chetyrkin:1997sg}
K.G.~Chetyrkin, B.A.~Kniehl, and M.~Steinhauser,
Phys.\ Rev.\ Lett.\  {\bf 79}, 2184 (1997);
Nucl.\ Phys.\ {\bf B510}, 61 (1998).

\bibitem{Schroder:2005hy}
Y.~Schr\"oder and M.~Steinhauser,
JHEP {\bf 0601}, 051 (2006);
K.G.~Chetyrkin, J.H.~K\"uhn, and C.~Sturm,
Nucl.\ Phys.\ {\bf B744}, 121 (2006).

\bibitem{Schroder:2005va}
Y.~Schroder and A.~Vuorinen,
JHEP {\bf 0506}, 051 (2005);
K.G.~Chetyrkin, M.~Faisst, C.~Sturm, and M.~Tentyukov,
Nucl.\ Phys.\ {\bf B742}, 208 (2006).

\bibitem{Chetyrkin:2006xg}
K.G.~Chetyrkin, J.H.~K\"uhn, C.~Sturm, 
Report No.\ SFB/CPP-06-21, TTP06-15, hep-ph/0604234;
R.~Boughezal, M.~Czakon, T.~Schutzmeier,
Report No.\ hep-ph/0605023;
R.~Boughezal and M.~Czakon,
Report No.\ hep-ph/0606232.

\bibitem{Kniehl:2005yc}
B.A.~Kniehl and A.V.~Kotikov,
Phys.\ Lett.\ B {\bf638}, 531 (2006);
Report No.\ DESY 06-073.

\bibitem{fur} W.~Furmanski and R.~Petronzio,
Z. Phys.\ C {\bf11}, 293 (1982).

\bibitem{Fleischer:1998nb}
J.~Fleischer, A.V.~Kotikov, and O.L.~Veretin,
Nucl.\ Phys.\ {\bf B547}, 343 (1999).

\bibitem{Melnikov:2000qh}
K.~Melnikov and T.~van Ritbergen,
Phys.\ Lett.\ B {\bf 482}, 99 (2000).

\end{thebibliography}
\end{document}